\documentclass[]{raa}            

\usepackage{graphicx,times}             
\usepackage{natbib}
\usepackage{amssymb,amsmath}
\bibpunct{(}{)}{;}{a}{}{,}
\usepackage[a4paper=true,dvipdfm=true,pagebackref=true]{hyperref}
\hypersetup{colorlinks = true, linkcolor = green, anchorcolor = red, citecolor = blue, filecolor = red, pagecolor = red, urlcolor = red}

\graphicspath{{img/}}
\usepackage{color}
\usepackage{ulem}

\usepackage[labelsep=space]{caption}
\usepackage{fancyhdr}

\fancypagestyle{cover}{%
	
	\lhead{}  
	\rhead{Apr. 9, 2018}
	\cfoot{}}

\begin{document}
\title{Mineralogy of northern nearside mare basalts
}

   \volnopage{Vol.0 (20xx) No.0, 000--000}      
   \setcounter{page}{1}          

   \author{Zhenchao Wang
      \inst{1,2}
   \and Yunzhao Wu
      \inst{3,2}$^{*}$\footnotetext{$^*$Corresponding author: Yunzhao Wu (wu@pmo.ac.cn)}
    \and Xiaomeng Zhang
      \inst{4}
    \and Yu Lu
      \inst{5,6}
   }

   \institute{Key Laboratory of Surficial Geochemistry, Ministry of Education, Department of Earth Sciences, Nanjing University, Nanjing 210023, China;
   	\and
   	Space Science Institute, Macau University of Science and Technology, Macau, China;
   	\and
   	Key Laboratory of Planetary Sciences, Purple Mountain Observatory, Chinese Academy of Sciences, Nanjing 210034, China;
   	\and
   	Xuzhou Bureau of Land and Resources, Xuzhou 221000, China;
   	\and
   	School of Geographic and Oceanographic Sciences, Nanjing University, Nanjing, 210023, China;
   	\and
   	Jiangsu Center for Collaborative Innovation in Geographical Information Resource Development and Application, Nanjing, 210023, China;\\
\vs\no
   {\small Received~~20xx month day; accepted~~20xx~~month day}\vspace{-5mm}}

\abstract{The mineralogy of mare basalts reflects the chemical 
composition of the magma source, as well as the physical and chemical 
environment of the rock's formation. It is significant for understanding the 
thermal evolution of the Moon. In this study, the spatial distribution of 
the mineralogy of the lunar northern nearside basalts was mapped using the 
Moon Mineralogy Mapper (M$ ^{3}) $ data. The study area, which is an elongated 
mare, Mare Frigoris and northern Mare Imbrium, was mapped and characterized 
into 27 units based on multi-source data, including spectrum, terrain, and 
element abundance. We extracted 177 M$ ^{3 } $ spectra from fresh craters. 
Spectral parameters such as absorption center and band area ratio (BAR) were 
obtained through data processing. The variation of mafic mineral of this 
region was acquired by analyzing these parameters. The basaltic units in eastern Mare Frigoris, which are older, have been found to be dominated by clinopyroxene with lower CaO compared to the returned lunar samples; this is similar to older 
basaltic units in Mare Imbrium. The basaltic units of western Mare Frigoris 
and Sinus Roris which are younger have been found to be rich in olivine. The late-stage basalts in 
Oceanus Procellarum and Mare Imbrium show the same feature. These 
wide-spread olivine-rich basalts suggest the uniqueness in the  
evolution of the Moon. Geographically speaking, Mare Frigoris is an 
individual mare, but the magma source region have connections with 
surrounding maria in consideration of mineral differences between western 
and eastern Frigoris, as well as mineral similarities with maria at the same 
location. \vspace{-3mm}
}

   \authorrunning{Z.-C. Wang, J. Chen, Y.-Z. Wu, X.-Y. Zhang, X.-M. Zhang \& Y. Lu, W. Cai}            
   \titlerunning{Mineralogy of northern nearside mare basalts}  

   \maketitle

%
%
\section{Introduction}           
\label{sect:intro}

Mare basalt covers 17{\%} of the surface of the Moon and accounts for 
approximately 1{\%} of the lunar crust (Head 1976). Its spatial 
distribution, composition, and volume record important information of the 
magmatic activity of the Moon. Research into its mineral components aids in the 
understanding of the magma source and characterizes the pressure, 
temperature, cooling rate, and other physical parameters of the magmatic 
state and processes. In addition, the mineralogy of mare basalts is an 
important scientific focus of the lunar exploration. 

Among the lunar maria, Mare Frigoris is very special. Most maria are circular; 
however, Mare Frigoris is very long and narrow (approximately 1800 km long 
with a maximum width of 200 km). Although Mare Frigoris is a separate mare 
geographically, its basalts may be associated with 
the adjacent maria (such as western Mare Frigoris  and Oceanus Procellarum, southern Mare Frigoris and northern  Mare Imbrium). Therefore, this study 
chose this long and narrow mare as the major research area to analyze the 
variation in mineral compositions in different locations. Further, it makes 
a comparison with the mineral compositions of adjacent maria (northern Mare 
Imbrium basalts) to discuss its role in the  thermal evolution of the 
Moon. 
\begin{figure}[!b]
	\begin{minipage}[t]{0.495\linewidth}
		\centering
		\includegraphics[width=\linewidth]{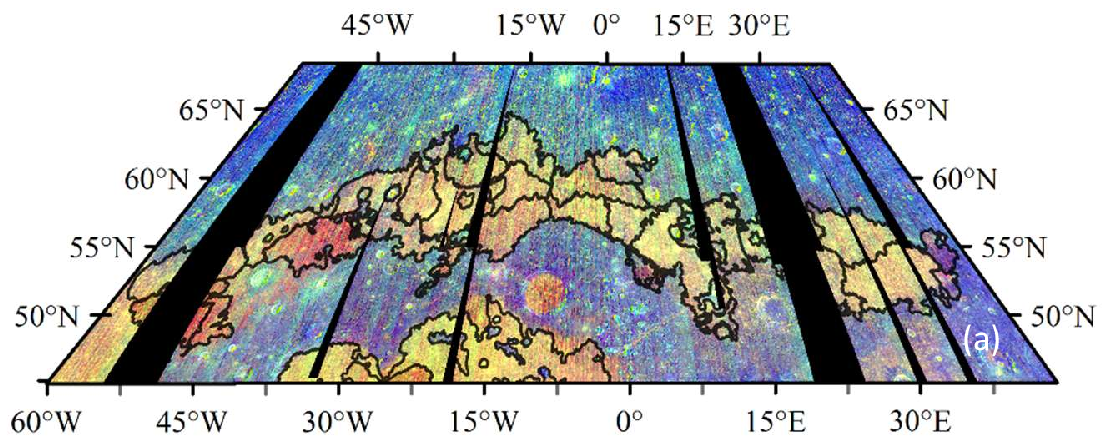}
	\end{minipage}\hfill
	\begin{minipage}[t]{0.495\textwidth}
		\centering
		\includegraphics[width=\linewidth]{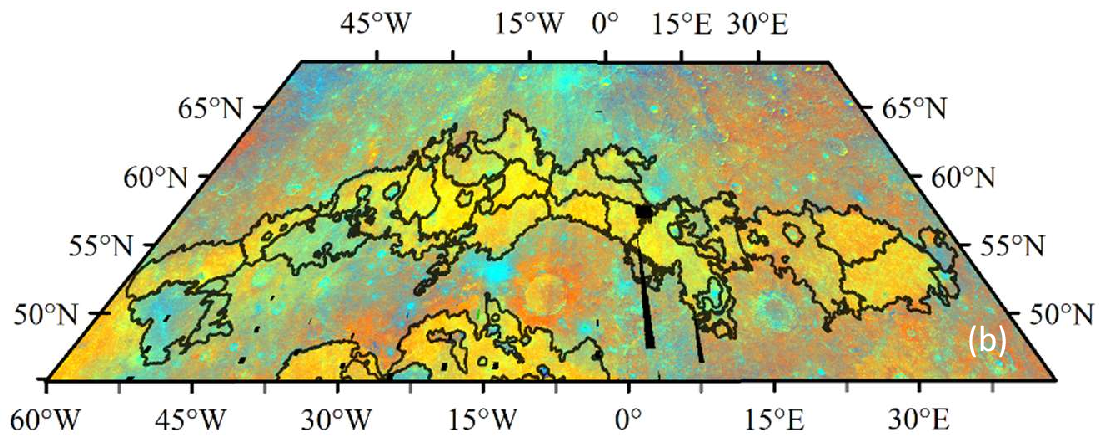}
	\end{minipage}\\
	\begin{minipage}[t]{0.495\linewidth}
		\centering
		\includegraphics[width=\linewidth]{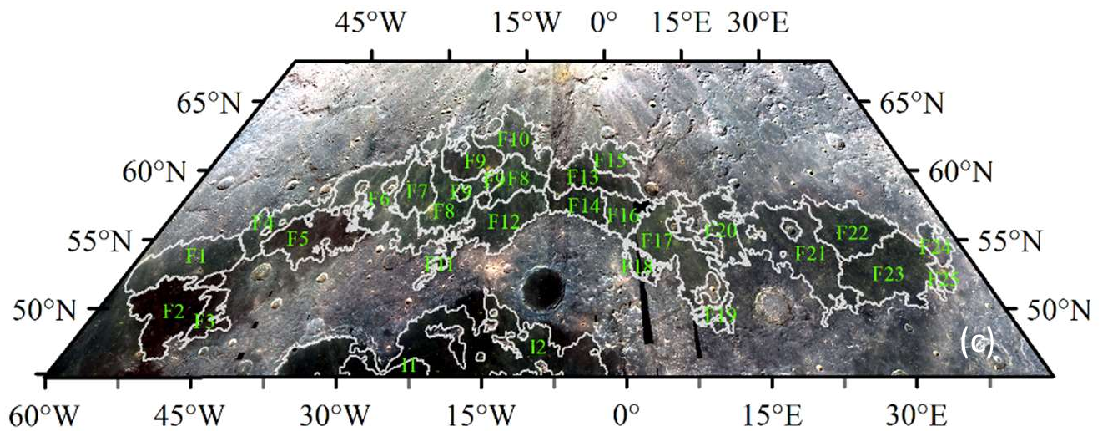}
	\end{minipage}\hfill
	\begin{minipage}[t]{0.495\textwidth}
		\centering
		\includegraphics[width=\linewidth]{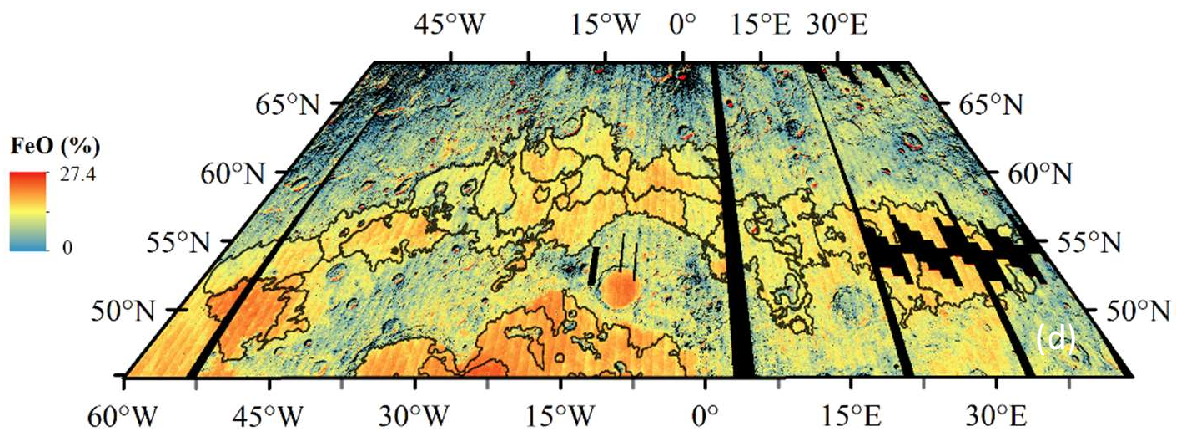}
	\end{minipage}\\
	\begin{minipage}[t]{0.495\linewidth}
		\centering
		\includegraphics[width=\linewidth]{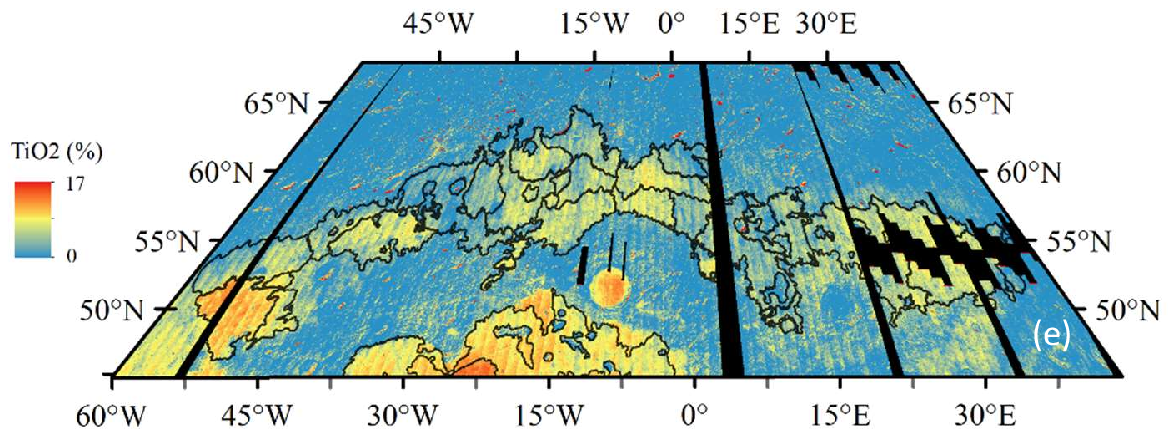}
	\end{minipage}\hfill
	\begin{minipage}[t]{0.495\textwidth}
		\centering
		\includegraphics[width=\linewidth]{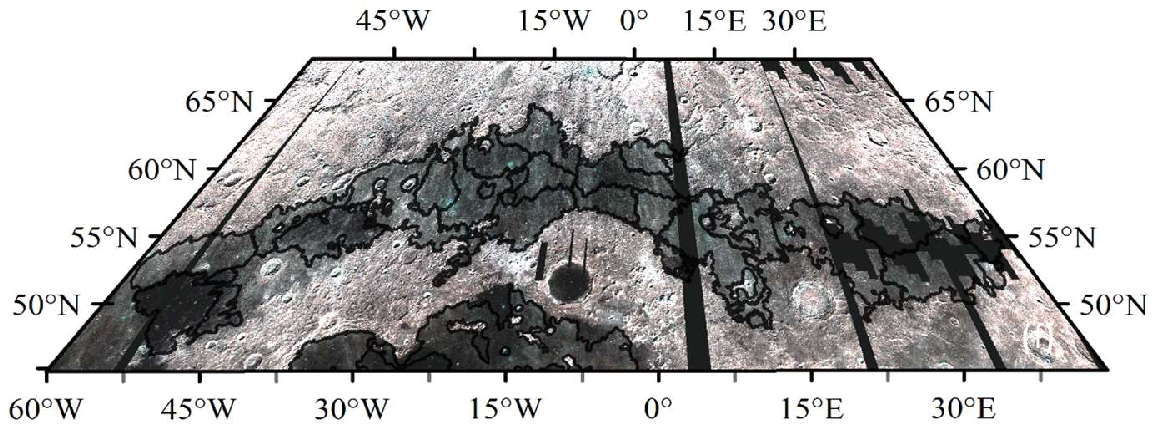}
	\end{minipage}%
	\caption{Images of the study area, where black/white lines delineate the 
		geological units divided in this study. (a) M$ ^{3} $ integrated band depth (IBD) image (R-1000 nm 
		IBD; G-2000 nm IBD; B-albedo of 1548 nm). (b) Clementine color composite 
		ratio image (R-750/415 nm, G-750/950 nm, B-415/750 nm). (c) Clementine false 
		color composite image (R-950 nm; G-750 nm; B-415 nm), in which F1-F25 
		represents  25 geological units of Mare Frigoris, and I1 and I2 represent 
		two geological units of Mare Imbrium. (d) The Chang'E-1 interference imaging spectroscopy (IIM) FeO content diagram. (e) IIM 
		TiO$ _{2} $ content diagram; (f) IIM false color composite image (R-891 nm; 
		G-739 nm; B-618 nm).}
	\label{fig1}
\end{figure}

Fig.~\ref{fig1} shows the focus area of this study. It centers around Mare Frigoris, 
including Sinus Roris in western and northern Mare Imbrium. Mare Frigoris is 
located in Procellarum KREEP Terrane (PKT) on the nearside of the Moon 
(Jolliff et al. 2000), at 56.0$ ^{\circ} $N central latitude and 
1.4$ ^{\circ} $E central longitude. Its west side is linked to Oceanus 
Procellarum by Sinus Roris. Whitford-Stark (1990) drew the geological map of 
Mare Frigoris based on ground observation and Lunar Orbiter data, and 
revealed that the oldest and the youngest basalts were located in the 
northeast and southwest, respectively. Hiesinger et al. (2010) divided 37 
geological units using the false color images of Clementine UV/VIS, and 
calculated the model age using the crater size-frequency distribution (CSFD) 
method. The results showed that the youngest basalts were located in Sinus 
Roris; four young Eratosthenian basalts were located in the middle of Mare 
Frigoris, northern Plato crater and northwest of Baily crater; the remaining 
old geological units were formed in the late Mare Imbrium, aged between 3.4 
and 3.8 Ga. Kramer et al. (2015) divided Mare Frigoris into 26 geological 
units based on Clementine, Lunar Prospector (LP), and Lunar Reconnaissance 
Orbiter (LRO) data. With the exception of two units in the west, where the 
contents of iron and titanium reached the levels of medium iron {\&} low 
titanium and medium iron {\&} medium titanium, respectively, the majority of 
the remaining areas were characterized as high-alumina basalts, 
characterized by low titanium to extremely low titanium, low iron, and high 
alumina, 

The Clementine UV/VIS data ranges from 415 to 1000 nm with only five bands. 
The spectral resolution is low, which restricts the reliable identification 
of the minerals. The India's Chandrayaan-1 Moon Mineralogy Mapper (M$ ^{3}) $ 
data have a wider range of wavelengths and higher spectral resolution. They 
can make an approximate description of the spectral curve continuously, so 
as to more accurately infer the mineral composition. The study obtains the 
spectral absorption characteristics of fresh craters by using of the M$ ^{3} $ 
data to identify the mineral composition of Mare Frigoris and northern Mare 
Imbrium. Then, it compares the composition with those of adjacent maria to 
discuss its regional geological significance.

\section{Data }
\label{sec2}

The M$ ^3 $ Level 2 (L2) data with spatial resolution of 140 m/pixel were used for the mineral analysis. In addition, this study adopts a variety of other data for the division of the stratigraphic units. Of these, the standard Clementine color ratio composite (R, 750/415, G, 750/950, B, 415/750) map and the standard M$ ^3 $ integrated band depth (IBD) composite (R: 1000 nm IBD; G: the 2000 nm IBD; and B: reflectance at 1480 nm) were used as base maps for separating basaltic units. In the two mosaic images the stitching traces between orbital boundaries are very obvious, which makes it difficult for the separation of different geologic units. The Lunar Reconnaissance Orbiter Camera (LROC) Wide Angle Camera (WAC) 7-channel composite (Wagner et al. 2015) and the Chang'E-1 interference imaging spectroscopy (IIM) data calibrated in Wu et al. (2013) eliminate the tile-shaped splicing boundary. These non-trace mosaics are conducive to the preparation  of the geological map, though they are partially missing and unable to completely cover the study area. Moreover, the M$ ^3 $ images containing thermal emission radiation, the WAC low-Sun mosaic and 30 m LOLA DEM show morphology information hence contributed to the separation of the basaltic units  (Wu et al. 2018a). The FeO and TiO$ _2 $ maps were from the original IIM 2C level data (Wu 2012). All the data except IIM were from the Planetary Data System (PDS).

\section{Methodology }
\label{sec3}

\subsection{Separation of Stratigraphy}

The stratigraphic division, which is important for the interpretation of mineralogy and geology, is not easy work because of the space weathering, contamination by the ejecta and lack of perfectly calibrated data. Following the methodology developed in Wu et al. (2018a), the separation of geologic units of this study was derived using diverse datasets including Clementine band ratio composite, FeO abundances, IIM color composite, LROC WAC mosaic composite, M$ ^3 $ IBD and M$ ^3 $ enhanced color, LROC WAC low-Sun image). All of these images were input in ArcGis and stretched for contrast enhancement. 

\subsection{Collection of Spectra}

To avoid space weathering, the spectra from small fresh craters with diameters of $ <\sim $1 km or slightly larger, and the mean spectra calculated from several pixels of the individual crater have been often used in previous publications (Kramer et al. 2008; Staid \& Pieters 2001; Whitten et al. 2011; Kaur et al. 2013). The extracted spectra from craters with diameters of $ \sim $1 km easily contain the compositions of the underlying stratigraphy because these craters of this size could excavate depths of approximately 84 m (Melosh 1989). If a crater has a lower resolution than that of M$ ^3 $, the spectra are not from fresh crater itself; instead, they contain shadows, melts, and soils. Through trial-and-error in Wu et al. (2018a), a fresh crater with a diameter of 3 -- 5 M$ ^3 $ pixels is optimal to not only distinguish walls and shadows, but also to represent the top unit. Following Wu et al. (2018a) in this study we used the fresh craters with diameters of 400 -- 500 m to sample spectra and they were from high-quality one pixel to reduce the pixel mixing. In order to ensure that the craters are fresh the very high  spatial-resolution LROC Narrow Angle Camera (NAC) data were used.

\subsection{Spectral parameter acquisition }

The mafic minerals (e.g., pyroxene and olivine) of mare basalts can be 
identified  through their characteristic spectral absorption features. 
Pyroxene displays two absorption peaks at approximately 1000 nm 
(Band I) and 2000 nm (Band II) (Adams 1974; 1975; Burns 1993; Cloutis 1985; 
Klima et al. 2007). In contrast, the olivine reflectance spectrum is 
revealed by a broad and asymmetric 1000 nm absorption, but lacks the 2000 nm absorption (Adams 1975; Singer 1981; Burns 1993; Isaacson et al. 2014). The 
broad Band I absorption in olivine is caused by three distinct absorption 
bands (Burns 1993). The central absorption, located just beyond 1000 nm, is 
caused by iron in the M2 crystallographic site. The two weaker absorptions 
near 850 and 1250 nm are the result of iron in the M1 site (Burns 1993; 
Sunshine {\&} Pieters 1998). The Band I ``secondary'' absorption near 1250 
nm allows olivine to be detected when admixed with the spectrally ``stronger'' pyroxene. 

The absorption band center is crucial for mineral 
identification. The Band I and Band II centers provide a qualitative way to 
access the composition of pyroxenes (Klima et al. 2007). The band centers 
are influenced by the amount of Fe$ ^{2+} $ and Ca$ ^{2+} $. For example, with 
increasing Fe$ ^{2+} $ and Ca$ ^{2+} $,  the band centers move slightly to the 
longer wavelength (Burns 1993; Hazen et al. 1978). However, in the case 
of olivine-pyroxene mixtures, Band I is dependent on the relative abundances 
of both olivine and pyroxene, which makes the interpretations problematic. 
The band area ratio (BAR), defined as the ratio between the Band II and Band 
I absorption features, is useful for estimating the relative abundances of 
pyroxene and olivine in basalts (Cloutis et al. 1986; Gaffey et al. 1993). The BAR is in inversely proportion to the olivine content, but increases linearly with pyroxene 
abundance (Gaffey et al. 1993; Cloutis et al. 1986).

Prior to calculating the absorption parameters, a B-spline function fitting 
is used to smoothen the spectral data to reduce the effects of spectral noise. 
The spectral curve after smoothing maintains a similar form and value to the 
original, with the noise removed. To derive the accurate absorption band center, the spectral continuum  should be removed. The continuum removal was performed using the convex 
hull method with the 2497 nm as the right endpoint suggested in Wu et al. 
(2018a) to handle the reflectance increases in the longer wavelength. The band 
area refers to the accumulative value of the absorption depths of the 
continuum function within 1000 and 2000 nm band absorption characteristic 
peaks. They are represented by Band I Area and Band II Area, respectively. 

\section{Results and Discussion}
\label{sec4}
\subsection{Stratigraphy}

According to the unit division method mentioned above, this study area has 
been divided into 27 unites, 25 units in Mare Frigoris and Sinus Roris and 
two units in the northern Mare Imbrium. Kramer et al. (2015) excludes F24 
from Mare Frigoris; however, in the LROC WAC and M$ ^{3} $ OP1B images, it is 
smooth and dark on the surface, and thus it is included in the scope of 
research. Owing to the serious loss of M$ ^{3} $ data around Lacus Mortis, it 
is impossible to extract the spectrum. Thus, it is not included in this 
study. Fig.~\ref{fig1} shows the unit division results in the M$ ^{3} $ IBD image, 
Clementine color ratio composite image, IIM FeO and TiO$ _{2} $ content 
diagram, and IIM false color composite image, respectively. 

Fig.~\ref{fig1}a shows the M$ ^3 $ (IBD) composite. In this figure, the highlands appear blue, whereas mare basalts  appear yellow/green to orange according to 
the relative strength of the mafic minerals. The darker red hues of the unit is, the 
higher olivine/pyroxene ratios (Staid et al. 2011). In Fig.~\ref{fig1}a, the F1, F4, and F6 units are in yellow, and F2, F3, and F5 are in 
orange. F7, F8, F9, and F10 are in yellow-green. F11 and F12 are in 
orange-red. F13 and F15 are in orange-yellow, and F14, F16, F17, and F19 are 
in yellow-green. Three edge units, namely, F18 in the southern Plato crater 
and F24 and F25 in the eastern Plato crater, are in purple in the Clementine 
ratio color composite image (Fig.~\ref{fig1}b), and, F18 is in red-purple and F24 is 
in blue-purple. The blue area in the Clementine ratio false color composite 
image indicates the high content of titanium in the unit.

\subsection{Spectral Analysis}

\begin{figure}[!b]
	\begin{minipage}[t]{0.495\linewidth}
		\centering
		\includegraphics[width=60mm]{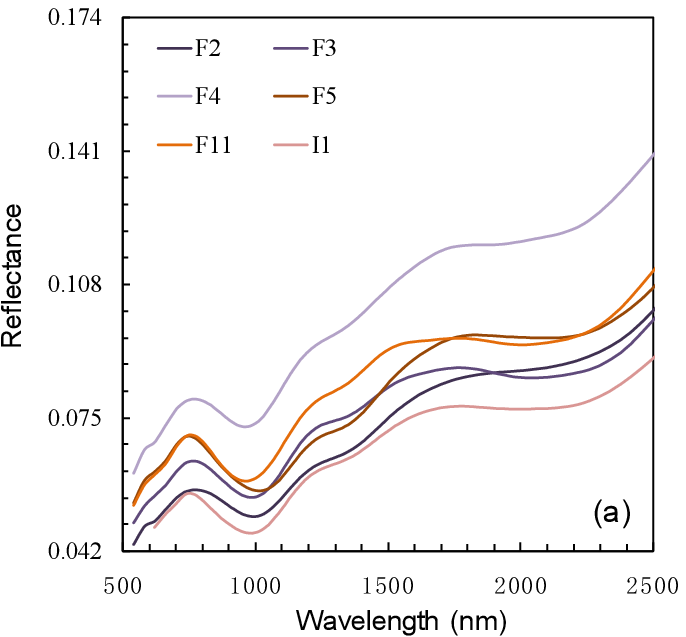}
	\end{minipage}\hfill
	\begin{minipage}[t]{0.495\textwidth}
		\centering
		\includegraphics[width=60mm]{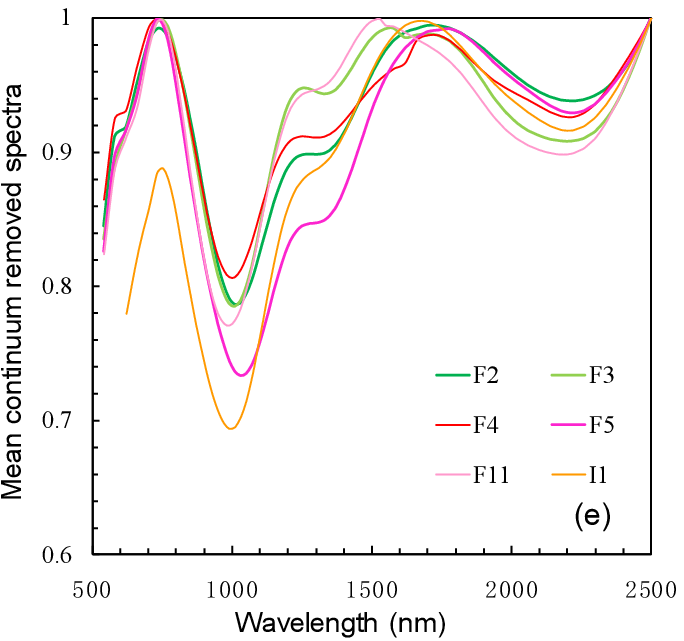}
	\end{minipage}\\
	\begin{minipage}[t]{0.495\linewidth}
		\centering
		\includegraphics[width=60mm]{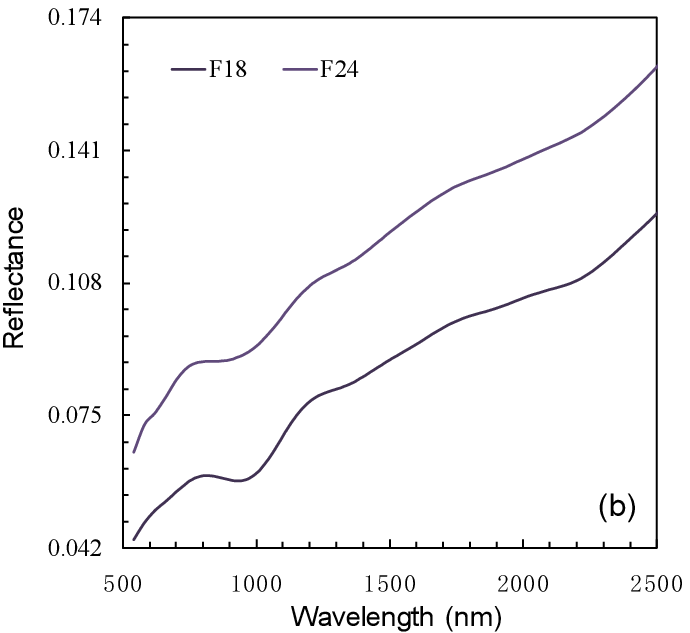}
	\end{minipage}\hfill
	\begin{minipage}[t]{0.495\textwidth}
		\centering
		\includegraphics[width=60mm]{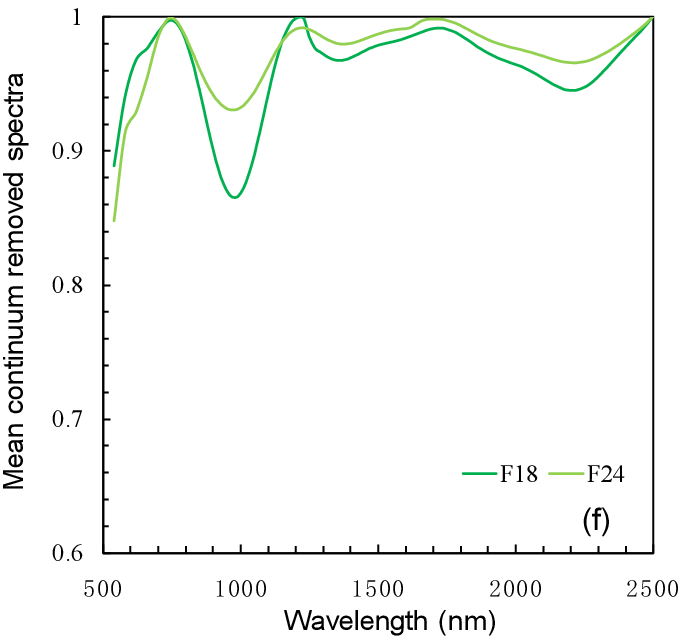}
	\end{minipage}\\
	\begin{minipage}[t]{0.495\linewidth}
		\centering
		\includegraphics[width=60mm]{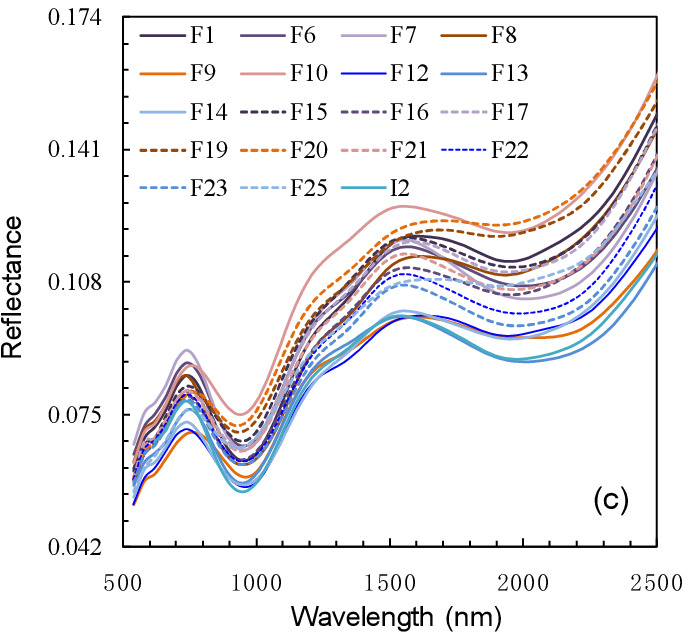}
	\end{minipage}\hfill
	\begin{minipage}[t]{0.495\textwidth}
		\centering
		\includegraphics[width=60mm]{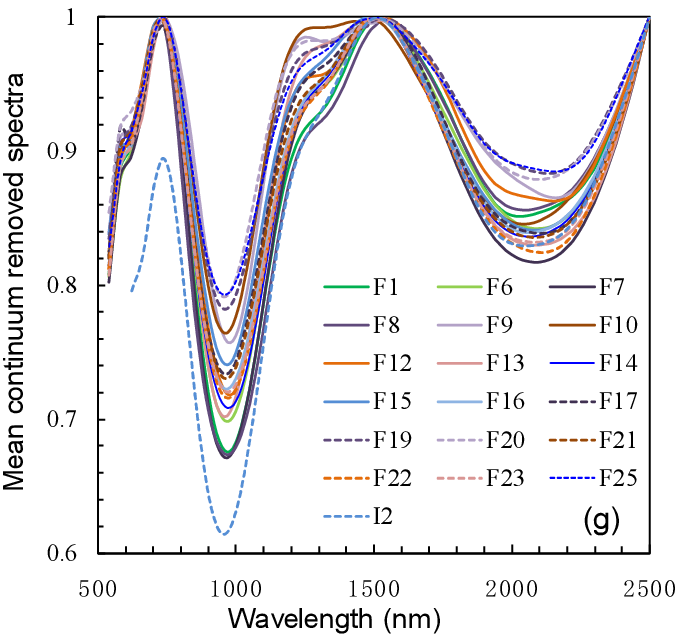}
	\end{minipage}%
	\caption{The average reflectance curves of the different group units and 
	continuum removed spectra of them. (a) G1: Strong absorption at 1000 nm and 
	weak absorption at 2000 nm; (b) G2: Weak absorption at both 1000 nm and 2000 
	nm; (c) G3: Strong absorption at both 1000 nm and 2000 nm. The continuum 
	removed spectra for (a), (b) and (c) are (e), (f) and (g) respectively.}
	\label{fig2}
\end{figure}

Fig.~\ref{fig2} shows the average reflectance of fresh craters, as well as their 
average reflectance curves after continuum removal. They are divided into 
three groups according to the shape of the spectral curves (width and depth 
of absorption characteristic and spectral slope): G1 (F2, F3, F4, and F5 in 
the western Mare Frigoris; F11 in the central Mare Frigoris; and I1 in the 
central northern Mare Imbrium), G2 (F18 in the southern Mare Frigoris, and 
F24 on the extreme eastern side of Mare Frigoris), and G3 (remaining units 
except those in G1 and G2). The spectra of most units have strong absorption 
characteristics at 1000 and 2000 nm. The absorption characteristic of five 
units in G1 at 1000 and 2000 nm are in the middle level of the three groups. 
They have an absorption at 1300 nm and weak absorption at 2000 nm. The 
absorption characteristic of two units in Group 2 is special; they have the 
weakest absorption at 1000 and 2000 nm and a big slope. This may be because 
it is difficult to select a fresh crater in this region hence contains the 
mixture of soils, or because the region has less crystallized composition. 
F24 is the oldest unit in Mare Frigoris, aged 3. 77 Ga (Hiesinger et al. 
2010). It has low of iron and titanium contents and a weak absorption 
characteristic of the spectral curve. The spectral curves of the units in G3 
are highly consistent; they have the strongest absorption at 1000 and 2000 
nm and weak absorption at 1300 nm. 
\begin{figure}[h]
	\begin{minipage}[t]{0.495\linewidth}
		\centering
		\includegraphics[width=60mm,height=60mm]{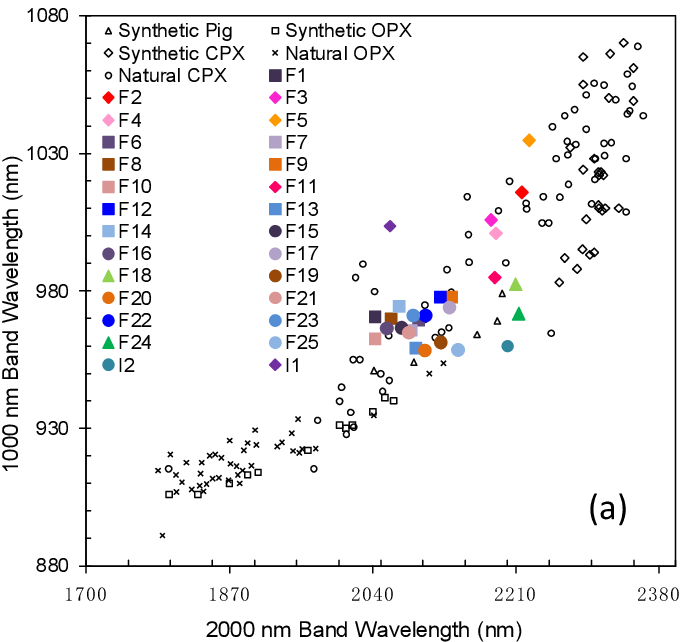}
	\end{minipage}\hfill
	\begin{minipage}[t]{0.495\textwidth}
		\centering
		\includegraphics[width=60mm,height=60mm]{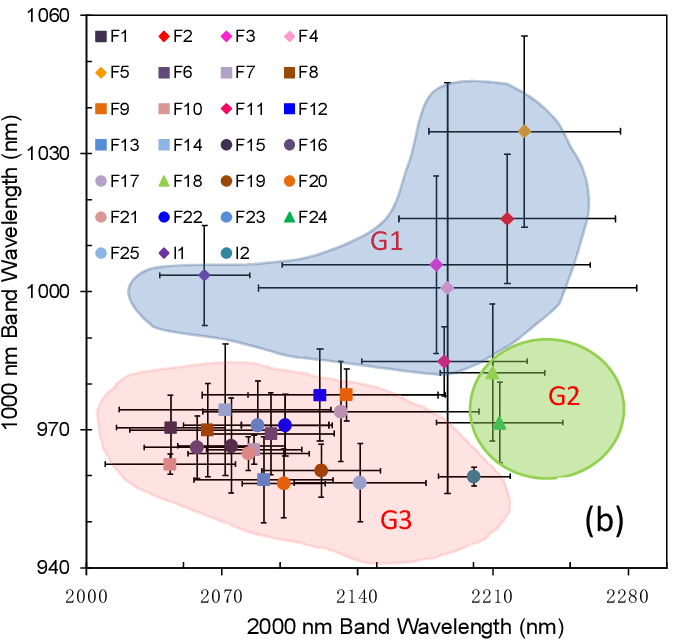}
	\end{minipage}%
	\caption{Plot of the 1000 nm (Band I) versus 2000 nm (Band II) absorption center. (a) Mean band center of each unit. The band centers of natural pyroxenes (Adams 1974; Cloutis \& Gaffey 1991) and synthetic pyroxene (Klima et al. 2011) are also included. (b) Mean band center with error bar of each measurement. The error bar represents the standard deviation of each measurement. The blue, green and pink lines circle G1, G2 and G3 group.}
	\label{fig3}
\end{figure}

Fig.~\ref{fig3} shows the average 1000 nm and 2000 nm absorption center scatter plots 
of all units. As a contrast, the figure also shows the laboratory spectral 
data of natural pyroxene and synthetic pyroxene (Adams 1974; Klima et al. 
2011). The wavelengths of the 1000 and 2000 nm absorption centers of 
synthetic low-Ca pyroxene, which better addresses the fundamental 
constraints of crystal structure and Ca-Mg-Fe content on reflectance 
spectra, are 900 to 930 nm and 1800 nm to 2100 nm, respectively. Those of 
synthetic high-Ca pyroxene are 950 nm to 1070 nm and 2260 nm to 2400 nm, 
respectively. G1 has the maximum absorption center wavelength, namely 985 nm 
to 1035 nm and 2181 nm to 2226 nm, respectively. The spectral curve has a 
wide and asymmetrical absorption peak at 1000 nm. The absorption peak is 
weak at 2000 nm, which is the olivine-dominated mineralogy. The absorption 
center wavelengths of G2 are 972 nm to 982 nm and 2210 nm to 2214 nm, 
respectively. Those of G3 are 958 nm to 978 nm and 2043 nm to 2141 nm, 
respectively, indicating rich monoclinic pyroxene; however, the content of 
pyroxene calcium is lower than that in the lunar samples. 

Fig.~\ref{fig3} shows the scatter plot of the absorption center wavelengths and BARs 
of the units of Mare Frigoris and northern Mare Imbrium. The BAR value is 
low in G1, ranging from 0.26 to 0.73, which is the olivine-dominated 
mineralogy. The BAR value is between 0.76 and 1.52 in G3, which is the 
clinopyroxenes and olivine mixed mineral. The BARs of F18 and F24 in G2 are 
0.85 and 0.83, respectively. 
\begin{figure}[!htb]\centering
	\includegraphics[width=0.5\linewidth]{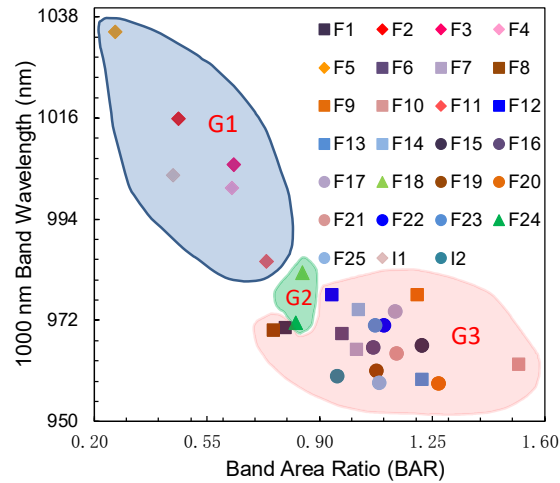}
	\caption{BAR versus Band I center plot. The blue, green and pink lines circle, G1, G2 and G3, the three groups of different units respectively.}
	\label{fig4}
\end{figure}

Fig.~\ref{fig4} presents the distribution diagram of the 1000 nm absorption centers 
and BAR values of Mare Frigoris and northern Mare Imbrium. Overall, they 
both show a certain negative correlation, which is in line with the 
proportional change in olivine/pyroxene. This indicates that the minerals 
comply with the change from olivine domination to pyroxene domination. The 
olivine-dominated mineralogy in western Mare Frigoris and central Mare 
Imbrium has a long 1000 nm absorption center wavelength and low BAR value. 
The clinopyroxenes-dominated mineralogy in eastern central Mare Frigoris has 
the shortest 1000 nm wavelength and a high BAR value. 
\begin{figure}[!htb]\centering
	\includegraphics[width=0.85\linewidth]{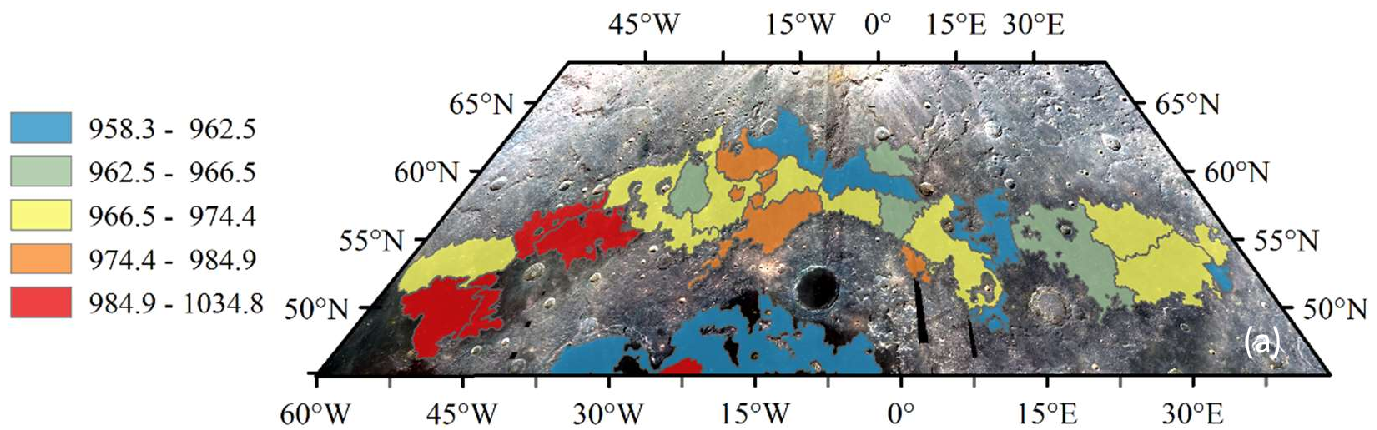}
	\includegraphics[width=0.85\linewidth]{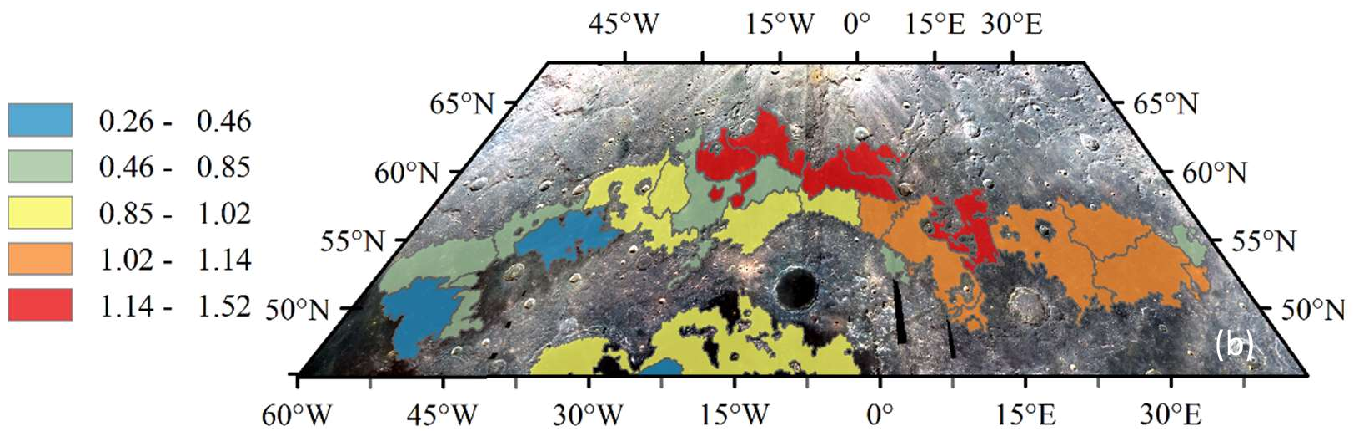}
	\caption{Distribution of Band I center (a) and BAR (b) values. The base map is the Clementine false color composite image (R-950 nm; G-750 nm; B-415 nm). }
	\label{fig5}
\end{figure}

The change in the relative trends of mafic mineral in space in Mare Frigoris 
and northern Mare Imbrium can be seen intuitively from Fig.~\ref{fig5}. Western Mare 
Frigoris, Sinus Roris and central Mare Imbrium (F2, F3, F4, F5, and I1) have 
the longest 1000 nm absorption center wavelengths and the smallest BAR, 
dominated by olivine. Young basalts in Oceanus Procellarum are also 
dominated by olivine (Staid et al. 2011; Zhang et al. 2016). Further 
combining with the age data, the basalts in western Mare Frigoris, Sinus 
Roris, Procellarum, and Mare Imbrium are all found to be young basalts (F2, 
F3, F4, and F5 aged 1.3-1.7 Ga (Hiesinger et al. 2010), and I1 aged 2.35 Ga 
(Wu et al. 2018a)). The similar olivine domination and age suggest that the 
late-stage thermal evolution of the Moon has a regularity in a larger scope.

The Mare Frigoris mineralogy in the northern Plato crater shows different 
absorption characteristic on the east and west sides. The 1000 nm absorption 
center on the west side is long with a small BAR value, whereas that on the 
east side is short with a large BAR value. The ratio of olivine to pyroxene 
on the west side is higher than that on the east. The 1000 nm absorption 
center wavelength and BAR value of the mineralogy around the eastern 
Aristoteles crater are 958.3 nm to 962.5 nm and 1.02 to 1.14, respectively. 
The ratio of olivine to pyroxene is low. The mineralogy in this area is old, 
aged over 3.56 Ga (Hiesinger et al. 2010). The basalts around the two 
craters are rich in low-Ca and medium-Ca pyroxene, which is consistent with 
the Mare Imbrium basalts on the north of the Mare Imbrium basin (Wu et al. 
2018a). This may indicate that the deep source of east central Mare Frigoris 
basalt has certain correlation with those of northern Mare Imbrium basalts. 
These phenomena indicate that the deep source of Mare Frigoris may be 
consistent with surrounding adjacent lunar maria, despite being a 
geographical mare.

\section{Summary and Conclusions}
\label{sec5}

By comprehensive utilization of multi-source data and according to various 
characteristics of albedo, hue, and topography, this study divided the units 
of the maria on the northern nearside of the Moon in detail, analyzed the 
spectral absorption characteristics of different units, mapped the 
distribution of the 1000 nm absorption center wavelengths and BAR values, 
and studied the mineral variation in different units.

The basalts in western Mare Frigoris and Sinus Roris have the highest ratio 
of olivine to pyroxene, i.e., they are rich in olivine. Similarly, the 
Eratosthenian basalts in northern Mare Imbrium (and Oceanus Procellarum) are 
also dominated by olivine  (Staid et al. 2011; Varatharajan et al. 2014; Zhang et al. 2016). The basalts in central and eastern Mare Frigoris 
have low ratios of olivine to pyroxene. The mineralogy with the lowest ratio 
is located around crater Fontenelle on the southern Mare Frigoris, and is 
rich in low-Ca to medium-Ca pyroxene. The old basalts in northern Mare 
Imbrium have similar composition. All of these phenomena indicate that, 
although Mare Frigoris is a geographical mare, the basalts  of Mare Frigoris 
have different magma sources and have a certain correlation to the sources 
of adjacent maria. 

The unsampled late-stage basalts are spectrally unique hence they are compositionally unique  and distinguished from older basalts. For the formation of these spectrally unique basalts some hypotheses have been suggested. One explanation is that the olivine-rich Eratosthenian basalts were formed via the partial melting from the ilmenite-rich mantle source and mixed with the residual KREEP layer beneath the anorthosite crust as it ascended to the surface (Zhang et al. 2015). Alternatively, Zhang et al. (2015) also suggests that the basalts were derived via the partial melting of the mantle source that mixed with  sinking ilmenite-rich KREEPy rocks. The young basalts in Mare Frigoris and Sinus Roris are rich in olivine, and the iron and titanium contents are also very high. The composition is similar to that of a large area of young basalts in Oceanus Procellarum and Mare Imbrium, which indicates that the late-stage volcanism of the Moon occurred on a larger scale. The young basalts have the lowest Mg\# among all the rocks of the Moon (e.g., Wu 2012; Crites \& Lucey 2015). The Chang'E-3 rover Active Particle-induced X-ray Spectrometer (APXS) data show that the young basalts are characterized with high-FeO (22.24 wt.\%), medium TiO\textsubscript{2} (4.31 wt.\%) and high Al\textsubscript{2}O\textsubscript{3} (12.11wt.\%)  (Wu et al. 2018b). The chemical and mineral data indicate that the young basalts do not have a KREEPy composition or affinity  (Neal et al. 2015). Therefore, such a large-scale evolution and the unique composition show that their sources have experienced a high evolution process  rather than assimilation with KREEPy materials. In the future, the samples of young basalts need to be collected for detailed petrological analysis. Chang'E-5 will collect samples, and it is recommended that the late-stage basalts be sampled.

\begin{acknowledgements}
	This research was supported by the National Key R\&D Program of China (2018YFB0504704), the National Natural Science Foundation of China (11773087), the Macau Science and Technology Development Fund (103/2017/A and 119/2017/A3) and Minor Planet Foundation of Purple Mountain Observatory. The M$ ^{3} $ Level 2 reflectance data are downloaded from the PDS (http: //pdsimaging.jpl.nasa.gov/volumes/m3.html). The LROC WAC images are acquired from the LROC website (http: //wms.lroc.asu.edu/lroc). The Clementine UV/VIS images are downloaded from the PDS note (http: //pdsimage.wr.usgs.gov).
\end{acknowledgements}

%
%
%


\label{lastpage}

\end{document}